# High thermoelectric performance of distorted Bismuth (110) layer

L. Cheng, H. J. Liu[*], J. Zhang, J. Wei, J. H. Liang, P. H. Jiang, D. D. Fan, L. Sun, J. Shi

*Key Laboratory of Artificial Micro- and Nano-Structures of Ministry of Education and School of Physics and Technology, Wuhan University, Wuhan 430072, China*

The thermoelectric properties of distorted bismuth (110) layer are investigated using first-principles calculations combined with the Boltzmann transport equation for both electrons and phonons. To accurately predict the electronic and transport properties, the quasiparticle corrections with the GW approximation of many-body effects have been explicitly included. It is found that a maximum *ZT* value of 6.4 can be achieved for *n*-type system, which is essentially stemmed from the weak scattering of electrons. Moreover, we demonstrate that the distorted Bi layer remains high *ZT* values at relatively broad regions of both temperature and carrier concentration. Our theoretical work emphasizes that the deformation potential constant characterizing the electron-phonon scattering strength is an important paradigm for searching high thermoelectric performance materials.

## 1. Introduction

Thermoelectric materials have attracted a lot of attention from the science community due to their interesting transport properties and potential applications in cooling and power generation. The efficiency of a thermoelectric material is quantified by the dimensionless figure of merit

$$ZT = \frac{S^2 \sigma T}{\kappa_e + \kappa_l} \tag{1}$$

where $S$ is the Seebeck coefficient, $\sigma$ is the electrical conductivity, $T$ is the absolute temperature, and $\kappa_e$ and $\kappa_l$ are the electronic and lattice thermal conductivity, respectively. In most thermoelectric materials, the transport coefficients $S$, $\sigma$, and $\kappa_e$ are interrelated in a way which makes it very challenging to achieve a high *ZT* value. Recently, several promising strategies have been proposed to improve the thermoelectric performance which include maximizing the power factor ($S^2\sigma$) through electronic doping and band engineering [1, 2], as well as minimizing the lattice thermal conductivity through phonon scattering [3]. Low-dimensional or nano-structured systems combine both advantages and thus are believed to exhibit significantly larger *ZT* values than their bulk counterparts [4, 5].

---

[*] Author to whom correspondence should be addressed. Electronic mail: phlhj@whu.edu.cn.

As a heavy element with strong spin-orbit coupling strength, bismuth is a common constitute of traditional thermoelectric materials and many well-known topological insulators [6, 7]. Previous theoretical work using first-principles method combined with Boltzmann theory found that bulk Bi has a maximum *ZT* value of 1.44 at 300 K [8]. It is reasonable to expect that Bi-based nanostructures could exhibit much higher thermoelectric performance. For example, it was theoretically predicted that the *ZT* value of Bi (111) monolayer could be optimized to as high as 4.1 [9]. On the experimental side, Nagao *et. al.* successfully grew the Bi (110) layers on the Si (111) surface and confirmed that it is more stable than the Bi (111) layers as long as the thickness is below four monolayers [10]. It is worth noting that the (110) layer has a distorted structure with atomic buckling, which can play an important role in determining its physical properties. A recent study also indicated that the distorted Bi (110) layer is more stable when the doping level is lower than 0.03 e/Å$^2$ [11]. It is therefore natural to ask whether the distorted Bi (110) layer possess good thermoelectric performance.

In this work, the thermoelectric properties of the distorted Bi (110) layer are investigated using first-principles calculations combined with Boltzmann transport theory for both electrons and phonons. The GW approximation is explicitly considered in order to accurately predict the electronic and transport properties. Due to very weak electron-phonon coupling, the distorted Bi (110) layer is found to exhibit high *ZT* values at relatively broad regions of both temperature and carrier concentration, which makes it a plausible candidate for high performance thermoelectric materials.

## 2. Computational Methods

Our density functional theory (DFT) calculations have been performed by using the projector augmented-wave method [12, 13] as implemented in the Vienna *ab initio* simulation package (VASP) code [14, 15, 16]. The local density approximation (LDA) is employed for the exchange and correlation functional [17]. The distorted Bi (110) layer is modeled by a rectangular lattice with four Bi atoms per unit cell, and the vacuum distance is set as 20 Å so that interactions between the layer and its periodic images can be treated as independent entities. For structural optimization, the energy cutoff is set as 400 eV and the Brillouin zone is sampled with 20×20×1 Monkhorst-Pack ***k*** meshes (corresponds to 100 irreducible ***k*** points). Optimal atom positions are determined until the magnitude of the force acting on each Bi atom become less than 0.01 eV/Å. The spin-orbit coupling (SOC) is explicitly considered in our calculations since Bi is heavy atom. For the GW calculations, we use a partially self-consistent GW0 in the present work, which means that our results are obtained by iterating only G, but keeping W fixed to the initial DFT W0. The SOC effect is also included at GW level which is fully relativistic for the core electrons and using the second-variation method for the valence states [18]. The Brillouin zone is sampled with 6×6×1 **Γ** centered grids (corresponds to 36 irreducible ***k*** points) after careful convergence test. The static dielectric function is evaluated with a sum over 180 unoccupied bands. For the maximally localized Wannier function generation process, we start with an initial guess of one $p_z$ and three $sp^2$ orbits per atom, and use the disentanglement technique to extract 32 Wannier functions. The electronic structures are then obtained using the Wannier interpolation formalism [19, 20, 21]. The electronic transport properties are computed using the semiclassical Boltzmann theory with the maximally localized Wannier function basis to interpolate the band structures and band velocities [22].

The lattice thermal conductivity is calculated by using the phonons Boltzmann transport equation as implemented in the ShengBTE code [23]. A 5×5×1 supercell is adopted to calculate the second-order and third-order interatomic force constant matrix. The interactions up to the fourth nearest neighbors are considered for the anharmonic one. For the three-phonon processes, a 30×30×1 **q**-point grid is needed to achieve convergence of thermal conductivity.

It should be emphasized that the SOC effect is included in the calculations of electronic structure at both LDA and GW level. It is, however, not included in the geometry optimization and frozen-phonon calculations. Such an approach has been widely used in the literatures [24, 25, 26]. In fact, it was previously shown that SOC has minor effects in the geometry optimization and phonon calculations [27, 28].

## 3. Results and Discussions

The atomic structure of the distorted Bi (110) layer is presented in Figure 1, where we see each Bi atom is covalently bonded with three neighboring Bi atoms to form a puckered two-dimensional hexagonal lattice. The structure is very similar to that of phosphorene [29], except that it has an additional sub-layer buckling $h = 0.6$ Å. It was previously found that such buckling makes the distorted Bi (110) layer more stable than the non-buckled one [30]. A recent study showed that the buckling is also closely related to the topological properties of Bi (110) layer [11]. As indicated in Fig. 1, the unit cell can be depicted by a rectangular lattice containing four Bi atoms with puckered surface. The optimized lattice parameters are $a = 4.38$ Å and $b = 4.55$ Å, and the Bi-Bi bond lengths are calculated to be 3.05 Å and 2.99 Å along the $x$ and $y$ directions, respectively. Compared with that of phosphorene having similar atomic structure, we find that the difference between lattice constant $a$ and $b$ in the distorted Bi (110) layer is much smaller, which can be attributed to the layer buckling since it tends to enlarge $a$ while decrease $b$.

Figure 2(a) and 2(b) show the phonon dispersion and the corresponding density of states (DOS) of the distorted Bi (110) layer. We see there is no imaginary frequency which indicates its kinetic stability. In fact, the distorted Bi (110) layer has been successfully grown on the Si (111) surface in ultrahigh vacuum [10]. The phonon band is nearly symmetric along the Γ-X and Γ-Y directions, which was also observed in the SnSe monolayer [31]. In contrast, the phonon dispersion of phosphorene has obvious asymmetry [32]. Such difference can be attributed to the additional sub-layer buckling in the Bi (110) and SnSe layers, while it disappears in the phosphorene. We further find that the group velocities of the longitudinal acoustic phonons are very small (2.50 km/s along the Γ-X direction and 2.25 km/s along the Γ-Y direction), which implies that the distorted Bi (110) layer may have a low lattice thermal conductivity $\kappa_l$. Fig. 2(c) plots the calculated $\kappa_l$ of Bi (110) layer as a function of temperature that ranges from 200 to 500 K. We do not consider higher temperatures since the melting point of Bi thin films are 520 K [33]. We see that $\kappa_l$ decreases with increasing temperature and approximately follows a 1/$T$ dependence, indicating that the Umklapp process is the dominant phonon scattering mechanism in the temperature range considered. At 300 K, the lattice thermal conductivity of Bi (110) layer is 4.9 W/mK and 6.3 W/mK along the *x* and *y* directions, respectively. Note these values are calculated with respect to a vacuum thickness of 6.6 Å (corresponds to interlayer distance in bulk Bi) and are lower than that of bulk counterpart (8 W/mK) [34], suggesting more favorable thermoelectric performance of Bi (110) layer.

We now move to the discussion of the electronic properties. Figure 3(a) shows the energy band structures of the distorted Bi (110) layer, where the calculations at LDA and GW levels are both shown. For the standard DFT calculations (referred to LDA in the figure), we find that the conduction band minimum (CBM) is located at the Γ-X direction, while the valence band maximum (VBM) is located at the Γ-Y direction. Moreover, there are several local band extrema a little higher (lower) than the CBM (VBM) along the Γ-Y, X-S and S-Y directions, which implies high DOS around the Fermi level (see Figure 3(b)). Detailed analysis of the orbital-decomposed DOS indicates that the 6*p* state of Bi plays leading role. As large DOS in the vicinity of Fermi level often leads to high Seebeck coefficients [35, 36], our observation suggests that the distorted Bi (110) layer may have large Seebeck coefficients for both *p*- and *n*-type systems. However, we should note that the CBM and/or VBM do not always locate at the high-symmetry lines for several thermoelectric materials [37, 38, 39, 40]. To check if this is also the case of our Bi (110) layer, we did a careful search in the Brillouin zone using a dense *k* mesh of 1000×1000×1. Indeed, we find that the real VBM is slightly off the Γ-Y direction and located at the **k**-point (0.030, 0.260, 0.000), which gives it a fourfold valley degeneracy and is beneficial to the thermoelectric performance. The CBM appears at the **k**-point (0.240, 0.000, 0.000), which is still located at the Γ-X high-symmetry line. As a result, the real band gap at the LDA level is calculated to be 0.17 eV. It should be noted that the precise calculations of the excited-state properties (such as band gap) are beyond the scope of traditional DFT. One approach to overcome this deficiency is to calculate the quasiparticle properties with the GW approximation of the many-body effects. As can be seen from Fig. 3, there is a downshift of the valance bands and an upshift of the conduction bands when the GW corrections are explicitly included in the calculations. Moreover, there is noticeable change of the band shape (especially for those along the Γ-Y direction), which would undoubtedly influence the band effective mass and thus the transport properties. There are also changes of the locations of CBM and VBM, which appear at the ***k*-**points of (0.248, 0.000, 0.000) and (0.014, 0.286, 0.000), respectively. As a result, the band gap at GW level is increased to 0.20 eV.

As GW calculations can give accurate band structures of Bi (110) layer, we will exclusively use it in the following. In combination with the Boltzmann transport theory, the electronic transport coefficients can be readily obtained. Within this approach, the electrical conductivity $\sigma$ can only be calculated with respect to the electron relaxation time $\tau$. Here $\tau$ is predicted by using the deformation potential (DP) theory [41, 42] considering the acoustic phonons are the main scattering mechanism. For a two-dimensional system, the relaxation time along the $\beta$ direction at temperature *T* can be expressed as

$$\tau_\beta = \frac{2\hbar^3 C_\beta}{3k_B T m_d^* E_\beta^2} \tag{2}$$

In this formula, $m_d^*$ is the density-of-state effective mass $m_d^* = (m_x^* m_y^*)^{1/2}$, where $m_x^*$ and $m_y^*$ are the effective mass along the *x* and *y* directions, respectively. $C_\beta$ is the elastic constant along the $\beta$ direction given by $C_\beta = \frac{1}{S_0} \frac{\partial^2 E}{\partial (\Delta l / l_0)^2}\bigg|_{l=l_0}$ ($E$ is the total energy of the system, $l_0$ is the lattice constant along the direction of $\beta$, $\Delta l = l - l_0$ is the corresponding lattice distortion, and $S_0$ is the equilibrium volume of the unit cell). The DP constant is calculated as $E_\beta = \frac{\partial E_{edge}}{\partial (\Delta l)/l_0}$, which represents the shift of band edges (VBM or CBM) per unit strain. These three quantities can be readily obtained from first-principles calculations. The other parameters $e$, $\hbar$, $k_B$ and $T$ are the unit charge, the reduced Planck constant, the Boltzmann constant, and the absolute temperature, respectively. The room temperature relaxation time of the distorted Bi (110) layer are summarized in Table I. Compared with those of conventional thermoelectric materials, we see that the relaxation time of the Bi (110) layer is one or two orders of magnitude larger for both electrons and holes, which is highly desirable for good thermoelectric performance. Moreover, we find that the electrons exhibit much higher relaxation time along the *y* directions, which is mainly attributed to very small DP constant (0.56 eV). Such a low value is a characterization of weak coupling of electrons and phonons, which would contribute to a better thermoelectric performance, as will be discussed in the following.

Figure 4(a)-(c) plots the room temperature Seebeck coefficient $S$, the electrical conductivity $\sigma$, the electronic thermal conductivity $\kappa_e$, and the power factor $S^2\sigma$ of the distorted Bi (110) layer as a function of carrier concentrations. It should be emphasized that for low-dimensional systems, the values of $\sigma$, $S^2\sigma$ and $\kappa_e$ depend on how to define the cross-sectional area or the vacuum region, which has some arbitrariness. In the present work, we use a uniform vacuum distance of 6.6 Å mentioned above when calculating both the electronic and phonon transport

coefficients so that the *ZT* value thus obtained does not depend on the arbitrary definition of vacuum region. For both the electrons and holes, we see from Fig. 4(a) that *S* exhibits peak values at smaller carrier concentrations. The absolute value can reach as high as 404 µV/K, which is much higher than those of conventional thermoelectric materials. Moreover, the absolute values of *S* are almost equal to each other for holes and electrons. In the single parabolic band model, *S* is proportional to the total density-of-states effective mass $m^* = N_v^{2/3} m_d^*$. Here $N_v$ is the above-mentioned valley degeneracy, which is 4 for VBM and 2 for CBM. The extra degeneracy compensates relatively lower $m_d^*$ of holes (0.110 $m_e$ compared with 0.173 $m_e$ of electrons) and eventually gives nearly identical $m^*$ for both carriers (0.277 $m_e$ of holes and 0.275 $m_e$ of electrons). As for the electrical conductivity, we see from Fig. 4(b) that there is a sharp increase of $\sigma$ for both holes and electrons as the carrier concentration is larger than $1.0 \times 10^{12}/cm^2$, which corresponds to the band edges of the distorted Bi (110) layer. In particular, $\sigma$ is significantly larger for electrons along the *y* directions, which is attributed to the much lower DP constants (0.56 eV) as mentioned above. It is well-known that the Seebeck coefficient and the electrical conductivity exhibit an opposite trend with the variation of carrier concentration. This means that there must be a trade-off between them so that a maximum power factor could be achieved at a particular carrier concentration, as can be seen from Fig. 4(c). As for the electronic thermal conductivity $\kappa_e$, we use the Wiedemann-Franz law $\kappa_e = L\sigma T$ where the Lorenz number *L* for a two-dimensional system can be expressed as

$$L = \frac{\kappa_e}{\sigma T} = \left(\frac{k_B}{e}\right)^2 \left[\frac{3F_2}{F_0} - \left(\frac{2F_1}{F_0}\right)^2\right] \quad (3)$$

with $F_i = F_i(\eta) = \int_0^\infty \frac{x^i dx}{e^{(x-\eta)}+1}$ ($\eta$ is the reduced Fermi energy). The calculated Lorenz number is $1.6\sim2.2\times10^{-8}$ V$^2$/K$^2$ in the carrier concentration ranges from $1.0\times10^{11}/cm^2$ to $1.0\times10^{14}$/cm$^2$. The electronic thermal conductivity thus calculated has virtually the same trend as the electronic conductivity (see Fig. 4(b)).

With all the transport coefficients discussed, we can now estimate the thermoelectric performance of the distorted Bi (110) layer. Fig. 4(d) plots the room temperature *ZT* value as a function of carrier concentration. In the case of *x* direction, a maximum *ZT* value of 2.9 and 2.5 can be achieved for the *p*- and *n*-type system, respectively. The corresponding carrier concentration is $1.7\times10^{12}$/cm$^2$ and $1.8\times10^{12}$/cm$^2$. For the *y* direction, the *ZT* value is 2.0 at a hole concentration of $2.2\times10^{12}$/cm$^2$, and 6.4 at an electron concentration of $5.8\times10^{11}$/cm$^2$. It should be mentioned that one can achieve the optimized carrier concentration by appropriate *n*-type and *p*-type doping. Taking *x* direction as an example, in Table II, we list the possible doping elements and optimal doping content for the distorted Bi (110) layer. On the other hand, a large *ZT* of *n*-type system is mainly attributed to its much higher electrical conductivity $\sigma$ as indicated by the significantly longer relaxation time $\tau$ and essentially stemmed from weaker electron-phonon coupling discussed above. Indeed, we see from Table I that the DP constant is much smaller (0.56 eV) for electrons compared with that of holes (3.00 eV). Our results emphasize that the weak electron-phonon coupling governed by smaller DP constant could contribute to better thermoelectric performance [43].

To discuss the temperature dependence of the thermoelectric performance, we plot in Figure 5 the optimized *ZT* values of the distorted Bi (110) layer as a function of temperature ranging from 200 K to 500 K. It can be seen that the *ZT* value of *y-n* system (*n*-type carriers along the *y* direction) reaches its maximum around 300 K and is much larger than those of *x-p*, *x-n*, and *y-p* systems over the whole temperature region. In contrast, the *ZT* values of *x-p*, *x-n*, and *y-p* increase with increasing temperatures, and all of them reach the maximum around 500 K. On the other hand, we see that the *ZT* value along the *x* direction is larger than that along the *y* direction for the *p*-type system, and it is just reversed for the *n*-type system. It should be mentioned that *ZT* values higher than 2.0 could be achieved for both *p*- and *n*-type systems at a relatively broad temperature range from 250 K to 500 K, which is very desirable for the practical applications of thermoelectric materials. Table III summarizes all the optimized *p*- and *n*-type *ZT* values of the distorted Bi (110) layer at different temperature. The corresponding carrier concentration, the transport coefficients, and the Lorenz number are also indicated. To further discuss the carrier concentration dependence of the thermoelectric performance, we list in Table IV the *ZT* values of distorted Bi (110) layer at a series of carrier concentration. At 300 K, we find that the *ZT* values are all higher than 2.0 when the carrier concentration is in the range from $1.0 \times 10^{12}/cm^2$ to $4.0 \times 10^{12}/cm^2$. At elevated temperature of 500 K, such range of carrier concentration becomes even larger ($2.0 \times 10^{12}/cm^2$ to $8.0 \times 10^{12}/cm^2$), suggesting big flexibility in the modulation of carrier concentrations to enhance the thermoelectric performance of the distorted Bi (110) layer.

## 4. Summary

In summary, our theoretical work demonstrates that the distorted Bi (110) layer could be optimized to exhibit a very high *ZT* of 6.4 at room temperature. Such value not only obviously exceeds that found in phosphorene with similar atomic structure (2.12) [44], but also outperforms that of Bi (111) layer (4.1) [9]. The significantly enhanced thermoelectric performance of the distorted Bi (110) layer is inherently connected with the weak electron-phonon coupling strength, as characterized by very small DP constant. Moreover, the *ZT* values are all higher than 2.0 at relatively broad regimes of both temperature and carrier concentration. To experimentally realize the strong prediction presented in this work, one needs to fabricate the Bi (110) single layer at first. In fact, Nagao *et al.* has successfully grown the distorted Bi (110) layers in an ultrahigh vacuum permalloy chamber with a thickness of about 1.5 nm [10]. Recently, Lu *et al.* deposited the Bi (110) single and double layers onto the highly oriented pyrolytic graphite [11]. With the rapid progress of fabrication techniques, it is reasonable to expect that high thermoelectric performance could be realized in the distorted Bi (110) layer, either with appropriate substrate or as a freestanding system.

## Acknowledgements

We thank financial support from the National Natural Science Foundation (Grant No. 11574236 and 51172167) and the "973 Program" of China (Grant No. 2013CB632502). We also acknowledge the financial support from the postgraduate programs under the "Fundamental Research Funds for the Central Universities" (Grant No. 2014202020206).

**Table I** The effective mass, DP constant, elastic constant, and relaxation time (at 300 K) of the distorted Bi (110) layer.

| Carrier type | $m_x^*(m_e)$ | $m_y^*(m_e)$ | $E_x$(eV) | $E_y$(eV) | $C_x$ (eV/Å$^2$) | $C_y$ (eV/Å$^2$) | $\tau_x$ (ps) | $\tau_y$ (ps) |
|---|---|---|---|---|---|---|---|---|
| hole | 0.076 | 0.158 | 4.80 | 3.00 | 2.79 | 1.52 | 0.143 | 0.126 |
| electron | 0.210 | 0.143 | 2.74 | 0.56 | 2.79 | 1.52 | 0.278 | 3.62 |

**Table II** Possible doping elements and optimal doping content for the distorted Bi (110) layer, which can be selected to maximize the thermoelectric performance along $x$ direction at 300 K. Results for $p$-type and $n$-type systems are both shown.

| *p*-type | | *n*-type | |
|---|---|---|---|
| element | content | element | content |
| Sn, C, Si, Pb | 0.00086 | S, Se, Te | 0.00087 |
| Ga, In | 0.00043 | Cl, Br, I | 0.00044 |

**Table III** Optimized *p*- and *n*-type *ZT* values of the distorted Bi (110) layer at different temperature. The corresponding carrier concentration, the transport coefficients, and the Lorenz number are also indicated.

| T (K) | | Concentration (10$^{12}$/cm$^2$) | S (μV/K) | σ (10$^6$/Ωm) | $S^2\sigma$ (W/mK$^2$) | L (10$^{-8}$ V$^2$/K$^2$) | $\kappa_e$ (W/mK) | $\kappa_l$ (W/mK) | ZT |
|---|---|---|---|---|---|---|---|---|---|
| 200 | x-p | 3.23 | 271 | 1.36 | 0.100 | 1.54 | 4.21 | 7.33 | 1.7 |
| | x-n | 2.69 | −223 | 1.36 | 0.068 | 1.59 | 4.33 | | 1.2 |
| | y-p | 3.46 | 249 | 1.30 | 0.081 | 1.56 | 4.06 | 9.36 | 1.2 |
| | y-n | 0.92 | −357 | 4.09 | 0.521 | 1.51 | 12.4 | | 4.8 |
| 250 | x-p | 2.40 | 293 | 1.09 | 0.094 | 1.54 | 4.21 | 5.88 | 2.3 |
| | x-n | 3.24 | −256 | 1.18 | 0.077 | 1.56 | 4.61 | | 1.8 |
| | y-p | 2.68 | 267 | 1.03 | 0.073 | 1.55 | 4.00 | 7.48 | 1.6 |
| | y-n | 0.62 | −375 | 3.79 | 0.533 | 1.50 | 14.2 | | 6.1 |
| 300 | x-p | 1.73 | 309 | 0.91 | 0.087 | 1.53 | 4.06 | 4.88 | 2.9 |
| | x-n | 1.76 | −277 | 1.06 | 0.081 | 1.54 | 4.91 | | 2.5 |
| | y-p | 2.08 | 278 | 0.91 | 0.070 | 1.54 | 4.21 | 6.27 | 2.0 |
| | y-n | 0.60 | −349 | 5.52 | 0.672 | 1.51 | 25.0 | | 6.4 |
| 350 | x-p | 1.26 | 317 | 0.82 | 0.082 | 1.52 | 4.36 | 4.18 | 3.4 |
| | x-n | 1.63 | −276 | 1.18 | 0.090 | 1.54 | 6.36 | | 3.0 |
| | y-p | 1.71 | 278 | 0.88 | 0.068 | 1.54 | 4.73 | 5.33 | 2.4 |
| | y-n | 0.62 | −323 | 7.94 | 0.828 | 1.52 | 42.1 | | 6.1 |
| 400 | x-p | 1.11 | 306 | 0.91 | 0.085 | 1.52 | 5.52 | 3.67 | 3.7 |
| | x-n | 1.66 | −265 | 1.36 | 0.096 | 1.55 | 8.45 | | 3.2 |
| | y-p | 1.68 | 266 | 1.00 | 0.071 | 1.55 | 6.21 | 4.67 | 2.6 |
| | y-n | 0.74 | −299 | 11.7 | 1.046 | 1.53 | 71.5 | | 5.5 |
| 450 | x-p | 1.05 | 292 | 1.06 | 0.090 | 1.53 | 7.30 | 3.27 | 3.8 |
| | x-n | 1.91 | −250 | 1.73 | 0.108 | 1.56 | 12.1 | | 3.2 |
| | y-p | 1.71 | 253 | 1.15 | 0.074 | 1.56 | 8.10 | 4.24 | 2.7 |
| | y-n | 0.89 | −281 | 16.2 | 1.279 | 1.54 | 112 | | 5.0 |
| 500 | x-p | 1.04 | 278 | 1.21 | 0.094 | 1.54 | 9.33 | 2.94 | 3.8 |
| | x-n | 2.26 | −237 | 2.12 | 0.119 | 1.57 | 16.7 | | 3.0 |
| | y-p | 1.89 | 239 | 1.33 | 0.076 | 1.57 | 10.5 | 3.73 | 2.7 |
| | y-n | 1.07 | −267 | 21.0 | 1.497 | 1.55 | 163 | | 4.5 |

**Table IV** The calculated *ZT* values at a series of carrier concentrations of the distorted Bi (110) layer. The results for two typical temperatures (300 and 500 K) are given for comparison.

| Temperature (K) | Carrier concentration ($\times 10^{12}/cm^2$) | *x* direction | | *y* direction | |
|---|---|---|---|---|---|
| | | holes | electrons | holes | electrons |
| 300 | 1.0 | 2.7 | 2.3 | 1.8 | 5.9 |
| | 2.0 | 2.9 | 2.5 | 2.1 | 4.3 |
| | 4.0 | 2.4 | 2.2 | 1.8 | 2.8 |
| | 8.0 | 1.5 | 1.5 | 1.2 | 1.6 |
| 500 | 1.0 | 1.1 | 0.9 | 0.8 | 1.7 |
| | 2.0 | 2.9 | 2.1 | 1.8 | 3.8 |
| | 4.0 | 3.8 | 3.0 | 2.7 | 4.4 |
| | 8.0 | 3.1 | 2.7 | 2.4 | 3.4 |

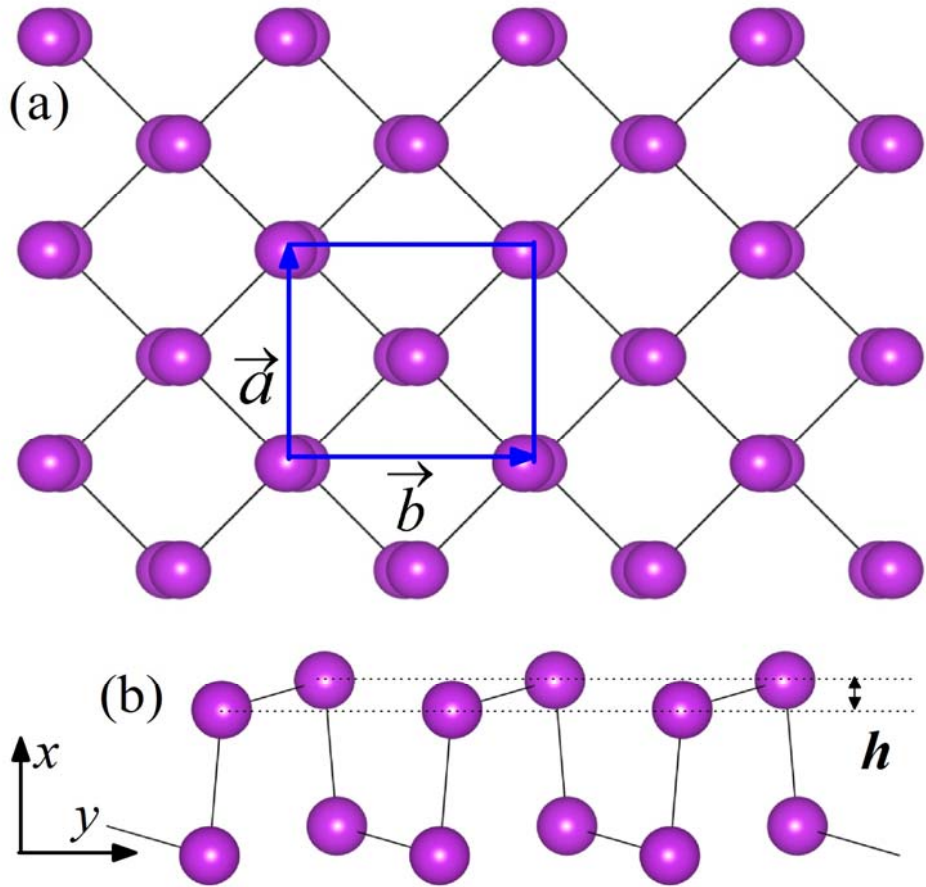

**Figure 1** (a) Top-, and (b) side- view of the distorted Bi (110) layer. The blue rectangle indicates the primitive cell, where $\vec{a}$ and $\vec{b}$ are the lattice vectors along the $x$ and $y$ directions, respectively. The layer buckling distance is indicated by $h$.

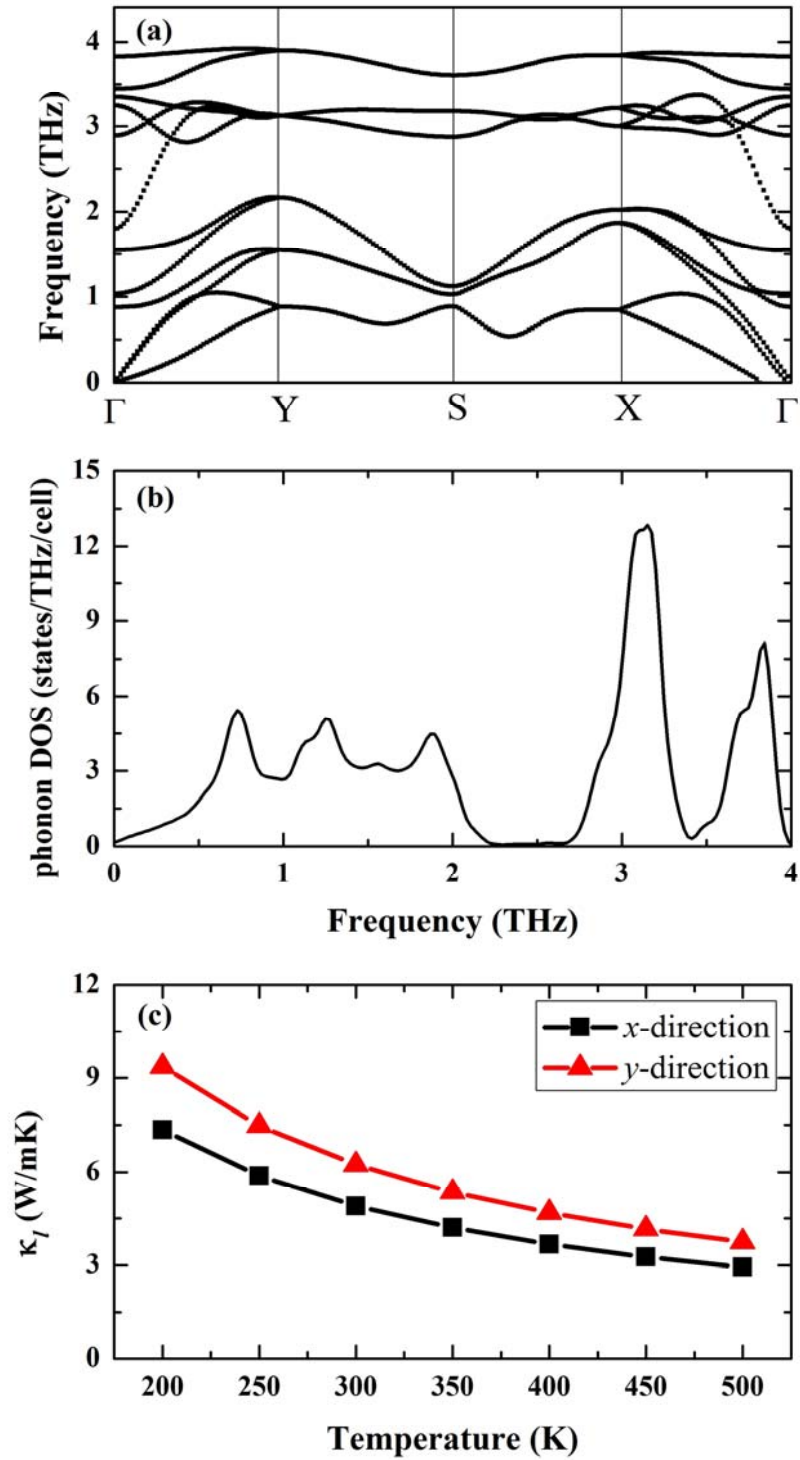

**Figure 2** (a) The phonon dispersion relations of the distorted Bi (110) layer, (b) the corresponding density of states, and (c) the temperature-dependent lattice thermal conductivity calculated with respect to a uniform vacuum thickness of 6.6 Å.

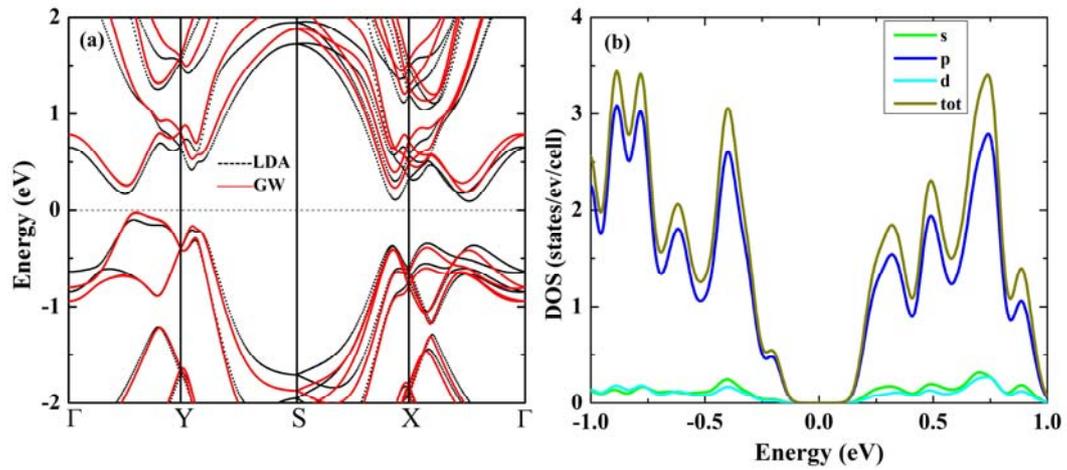

**Figure 3** (a) The energy band structures of the distorted Bi (110) layer, where calculations at the LDA and GW levels are both shown. (b) plots the density of states at the LDA level, where the contributions from s, p, and d states are also shown. The Fermi level is at 0 eV.

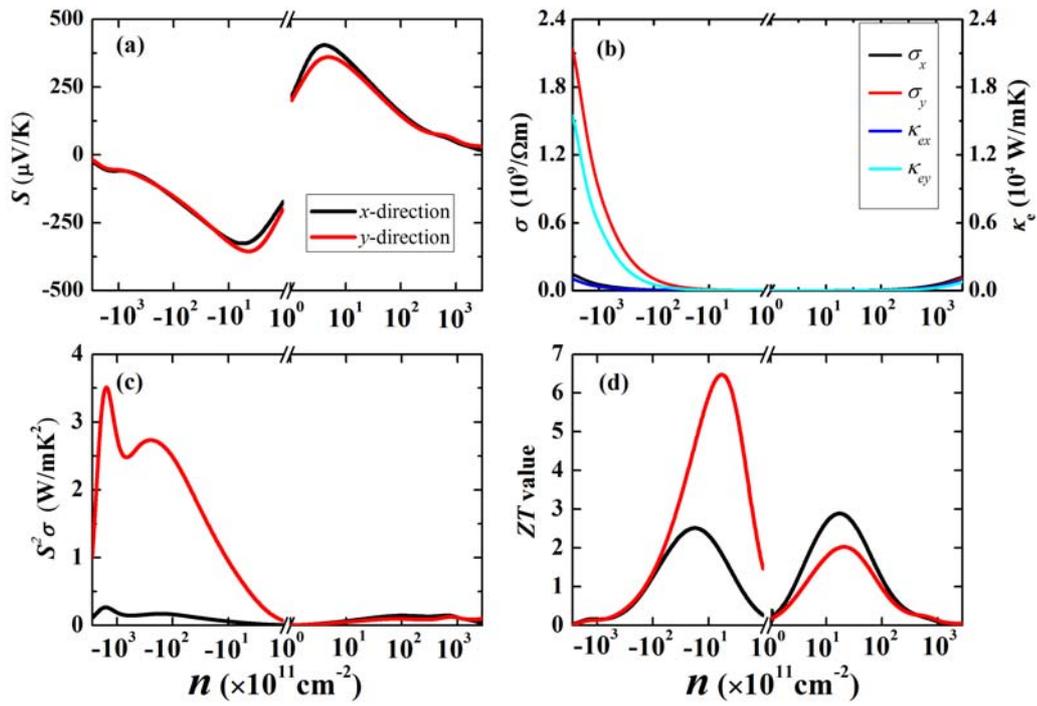

**Figure 4** The room temperature (a) Seebeck coefficient $S$, (b) electrical conductivity $\sigma$ and electronic thermal conductivity $\kappa_e$, (c) power factor $S^2\sigma$, and (d) $ZT$ value of the distorted Bi (110) layer as a function of carrier concentration along the $x$ and $y$ directions. Positive and negative carrier concentrations represent $p$- and $n$-type carriers, respectively. The transport coefficients are renormalized with respect to a uniform vacuum thickness of 6.6 Å.

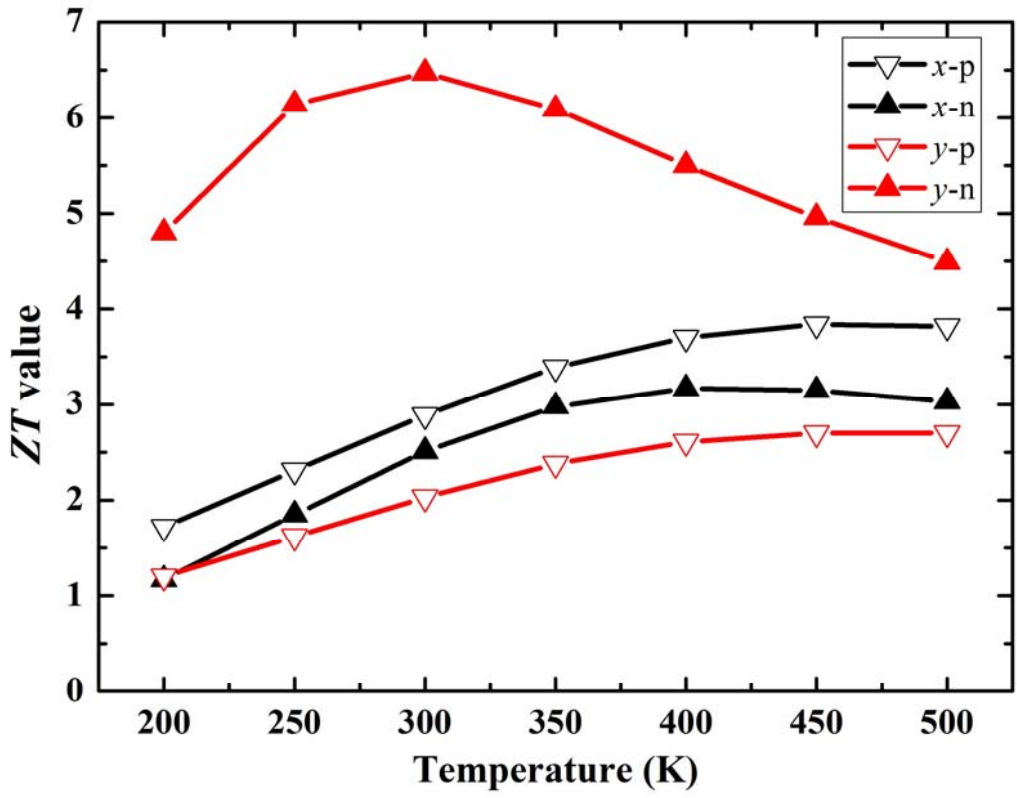

**Figure 5** The temperature dependence of the *ZT* values of the distorted Bi (110) layer.